\documentclass[fleqn,twoside]{article}
\usepackage{espcrc2}

\newcommand{\AmS}{{\protect\the\textfont2
  A\kern-.1667em\lower.5ex\hbox{M}\kern-.125emS}}

\title{Design of modular wireless sensor}

\author{Mark Sh. Levin
\thanks{
 Mark Sh. Levin:~
 Inst. for Inform. Transmission Problems,
 Russian Academy of Sciences, 19 Bolshoj Karetny lane, Moscow 127994, Russia.
 Email: mslevin@acm.org
  }
 and
 Alexander V. Fimin
 \thanks{
 Alexander V. Fimin:~
 NetCracker Technology,
 83 Dubninskaya Str.,
 Moscow 127591, Russia.
 Email: fimin@frtk.ru
 }
  }

\begin{document}

\begin{abstract}
 The paper addresses
 combinatorial approach to
  design of modular wireless sensor
 as composition of the sensor element
 from its component alternatives
 and aggregation of the obtained solutions into a resultant
 aggregated solution.
 A hierarchical model is used for
 the wireless sensor element.
 The solving process consists of three stages:
%
%
 (i) multicriteria ranking
  of design alternatives for system components/parts,
 (ii) composing the selected design alternatives
 into composite solution(s)
 while taking into
 account ordinal quality of the design alternatives above
 and their compatibility (this stage is based on
 Hierarchical Morphological Multicriteria Design
 - HMMD), and
 (iii) aggregation of the
 obtained composite solutions into a resultant
 aggregated solution(s).
 A numerical example
 describes the problem structuring and solving processes
 for modular alarm wireless sensor element.

~~


\vspace{1pc}
\end{abstract}

\maketitle

\newcounter{cms}
\setlength{\unitlength}{1mm}

\section{Introduction}

 In recent years the significance of sensor systems/networks is increased
 (e.g., \cite{aky02},
   \cite{cul04},
  \cite{kul11},
 \cite{qi01},
  \cite{soh07},
 \cite{zad07}).
 In general,
 it may be reasonable to consider a simplified 3-layer architecture of
 a
 sensor system
 (Fig. 1):
 (i) sensors and sensor local networks (sensor subsystem layer),
 (ii) communication network (transportation layer), and
 (iii) management subsystem (
 control layer: information analysis and integration/fusion,
 decision making and control)
  (e.g.,
   \cite{kul11},
  \cite{qi01},
  \cite{zad07}).
 In the article, a hierarchical modular design of
 configuration for wireless sensor element is examined.
  The problem corresponds to  layer 1
 of the three-layer sensor system structure above.
 A real world numerical example is targeted
  to a fire alarm wireless sensor element.
 Note various approaches have been applied for the design of system configurations
 \cite{lev09}:
 (1) the shortest path problem \cite{art91};
  (2) evolutionary approaches
  (e.g.,
  \cite{rod05});
%
 (3) multi-agent approaches (e.g.,
  \cite{campbell03});
 (4) approaches based on fuzzy sets
 (e.g., \cite{smirnov04});
  (5)
 composite constraint satisfaction problems
 (e.g.,
 \cite{sabin98}, \cite{stefik95});

\begin{center}
\begin{picture}(70,71)
\put(04,00){\makebox(0,0)[bl]{Fig. 1. Sensor system (three layers)
\cite{levfim10}}}


\put(03,46){\line(1,0){64}} \put(03,69){\line(1,0){64}}
\put(03,46){\line(0,1){23}} \put(67,46){\line(0,1){23}}

\put(03.5,46.5){\line(1,0){63}} \put(03.5,68.5){\line(1,0){63}}
\put(03.5,46.5){\line(0,1){22}} \put(66.5,46.5){\line(0,1){22}}

\put(6.5,64){\makebox(0,0)[bl]{Layer 3: Management }}

\put(6.5,60){\makebox(0,0)[bl]{(monitoring processes,
information}}

\put(6.5,56){\makebox(0,0)[bl]{analysis and processing/fusion,}}
\put(6.5,52){\makebox(0,0)[bl]{information collection and store,}}
\put(6.5,48){\makebox(0,0)[bl]{generation of control decisions)}}


\put(10,41){\vector(1,1){5}}

\put(22.5,41){\vector(0,1){5}}


\put(35,41){\vector(0,1){5}}

\put(47.5,41){\vector(0,1){5}}


\put(60,41){\vector(-1,1){5}}


\put(35,31){\oval(70,20)}

\put(5,36){\makebox(0,0)[bl]{Layer 2: Transportation
(information}}

\put(5,32){\makebox(0,0)[bl]{transmission and processing,
routing,}}

\put(5,28){\makebox(0,0)[bl]{scheduling, network
covering/spanning, }}

\put(5,24){\makebox(0,0)[bl]{preliminary data aggregation)}}


\put(05,16){\vector(0,1){5}} \put(15,16){\vector(0,1){5}}
\put(25,16){\vector(0,1){5}} \put(35,16){\vector(0,1){5}}

\put(45,16){\vector(0,1){5}} \put(55,16){\vector(0,1){5}}
\put(65,16){\vector(0,1){5}}

\put(00,06){\line(1,0){70}} \put(00,16){\line(1,0){70}}
\put(00,06){\line(0,1){10}} \put(70,06){\line(0,1){10}}

\put(4,11.5){\makebox(0,0)[bl]{Layer 1: Sensors (including
preliminary}}

\put(4,07.5){\makebox(0,0)[bl]{data processing and transmission)}}

\end{picture}
\end{center}

 (6) ontology-based approaches
 (e.g.,
 \cite{ciuksys07});
%
 (7) multicriteria multiple choice problem
 (e.g.,
 \cite{poladian06});
 (8) hierarchical multicriteria morphological design (HMMD) approach
 (\cite{lev98},
 \cite{lev06},
 \cite{lev09});
%
%
 (9) AI techniques
 (e.g.,
 \cite{mcd82},
 \cite{pira05},
 \cite{sugu08},
 \cite{wielinga97});
 and
 (10) design grammars approaches
 (e.g., multidisciplinary grammar approach that includes
 production rules and optimization, graph grammar approach)
 (e.g.,
 \cite{schmidt00}).
%
%
 A survey of combinatorial optimization approaches to system configuration design
 is presented in
  \cite{lev09}.

 In this article,
 a generalized composite design framework
 for modular systems
 is used (Fig. 2):
 selection of design alternatives (DAs) for system
 components/parts,
 combinatorial synthesis (composition) of the composite
 solutions, and
 aggregation of the obtained solutions to get a resultant
 aggregated solution.

%

\begin{center}
\begin{picture}(72,79)

\put(02,00){\makebox(0,0)[bl]{Fig. 2.
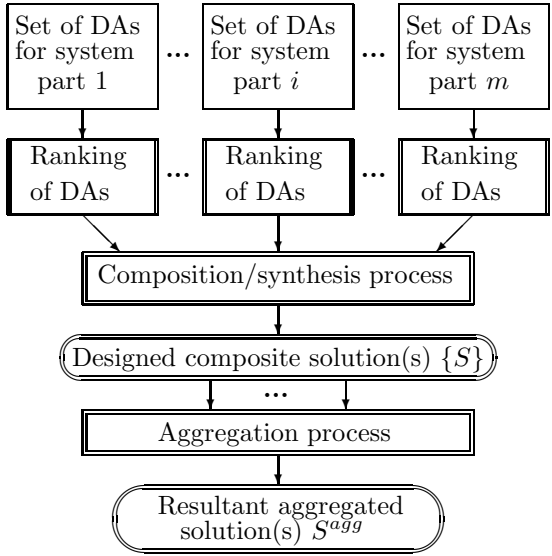
 Selection, composition, aggregation
}}


\put(00,64){\line(1,0){20}} \put(00,78){\line(1,0){20}}
\put(00,64){\line(0,1){14}} \put(20,64){\line(0,1){14}}

\put(01,74){\makebox(0,0)[bl]{Set of DAs}}
\put(01,70){\makebox(0,0)[bl]{for system}}
\put(04,66){\makebox(0,0)[bl]{part \(1\)}}

\put(10,64){\vector(0,-1){4}}


\put(21,71){\makebox(0,0)[bl]{{\bf ...}}}
\put(47,71){\makebox(0,0)[bl]{{\bf ...}}}


\put(26,64){\line(1,0){20}} \put(26,78){\line(1,0){20}}
\put(26,64){\line(0,1){14}} \put(46,64){\line(0,1){14}}

\put(27,74){\makebox(0,0)[bl]{Set of DAs}}
\put(27,70){\makebox(0,0)[bl]{for system }}
\put(30,66){\makebox(0,0)[bl]{part \(i\)}}

\put(36,64){\vector(0,-1){4}}


\put(52,64){\line(1,0){20}} \put(52,78){\line(1,0){20}}
\put(52,64){\line(0,1){14}} \put(72,64){\line(0,1){14}}

\put(53,74){\makebox(0,0)[bl]{Set of DAs}}
\put(53,70){\makebox(0,0)[bl]{for system }}
\put(56,66){\makebox(0,0)[bl]{part \(m\)}}

\put(62,64){\vector(0,-1){4}}



\put(00,50){\line(1,0){20}} \put(00,60){\line(1,0){20}}
\put(00,50){\line(0,1){10}} \put(20,50){\line(0,1){10}}

\put(0.4,50){\line(0,1){10}} \put(19.6,50){\line(0,1){10}}

\put(03,56){\makebox(0,0)[bl]{Ranking}}
\put(03,52){\makebox(0,0)[bl]{of DAs }}

\put(10,50){\vector(1,-1){5}}


\put(21,55){\makebox(0,0)[bl]{{\bf ...}}}
\put(47,55){\makebox(0,0)[bl]{{\bf ...}}}


\put(26,50){\line(1,0){20}} \put(26,60){\line(1,0){20}}
\put(26,50){\line(0,1){10}} \put(46,50){\line(0,1){10}}

\put(26.4,50){\line(0,1){10}} \put(45.6,50){\line(0,1){10}}

\put(29,56){\makebox(0,0)[bl]{Ranking}}
\put(29,52){\makebox(0,0)[bl]{of DAs  }}

\put(36,50){\vector(0,-1){5}}


\put(52,50){\line(1,0){20}} \put(52,60){\line(1,0){20}}
\put(52,50){\line(0,1){10}} \put(72,50){\line(0,1){10}}

\put(52.4,50){\line(0,1){10}} \put(71.6,50){\line(0,1){10}}

\put(55,56){\makebox(0,0)[bl]{Ranking}}
\put(55,52){\makebox(0,0)[bl]{of DAs }}

\put(62,50){\vector(-1,-1){5}}


\put(10,38){\line(1,0){52}} \put(10,45){\line(1,0){52}}
\put(10,38){\line(0,1){07}} \put(62,38){\line(0,1){07}}

\put(10.5,38.5){\line(1,0){51}} \put(10.5,44.5){\line(1,0){51}}
\put(10.5,38.5){\line(0,1){06}} \put(61.5,38.5){\line(0,1){06}}

\put(12,40){\makebox(0,0)[bl]{Composition/synthesis process}}

\put(36,38){\vector(0,-1){04}}


\put(36,31){\oval(58,6)} \put(36,31){\oval(57,5)}

\put(08.6,29){\makebox(0,0)[bl]{Designed composite solution(s)
\(\{S\}\)}}


\put(27,28){\vector(0,-1){04}}

\put(34,25.7){\makebox(0,0)[bl]{{\bf ...}}}

\put(45,28){\vector(0,-1){04}}


\put(10,18){\line(1,0){52}} \put(10,24){\line(1,0){52}}
\put(10,18){\line(0,1){06}} \put(62,18){\line(0,1){06}}

\put(10.5,18.5){\line(1,0){51}} \put(10.5,23.5){\line(1,0){51}}
\put(10.5,18.5){\line(0,1){05}} \put(61.5,18.5){\line(0,1){05}}

\put(20,19.2){\makebox(0,0)[bl]{Aggregation process}}

\put(36,18){\vector(0,-1){04}}


\put(36,09.5){\oval(44,9)} \put(36,9.5){\oval(43,8)}

\put(20,9.5){\makebox(0,0)[bl]{Resultant aggregated }}

\put(23,6){\makebox(0,0)[bl]{solution(s) \(S^{agg}\)}}

\end{picture}
\end{center}

 The approach is based on three optimization problems:

 {\bf I.} Multicriteria ranking
 (outranking technique as a modification of ELECTRE method
 is used
  \cite{roy96}).

 {\bf II.} Morphological synthesis based on morphological clique problem
 (as Hierarchical Morphological Multicriteria Design - HMMD)
 (\cite{lev98},
   \cite{lev06},
  \cite{lev09}, \cite{lev12}).
 {\bf III.} Aggregation of the obtained composite solutions into
 the
 resultant
 aggregated solution(s)
 (aggregation strategies are used,
 e.g.,  design of system ``kernel'' and its extension)
 \cite{lev11agg}
 (here knapsack-like problems are sued).


%
 The illustrative numerical design example involves
 hierarchical structure of sensor
 (and-or tree model),
 design alternatives (DAs) for system parts/components,
 Bottom-Up solving process.
 Estimates of DAs and their compatibilities are based on expert judgment.

 A preliminary material of the paper was published as conference paper
 \cite{levfim10}.


\section{Structure of Sensor and Estimates}

 The following simplified illustrative hierarchical structure of an alarm wireless sensor
 element is examined
 (Fig. 3):

\begin{center}
\begin{picture}(74,96)
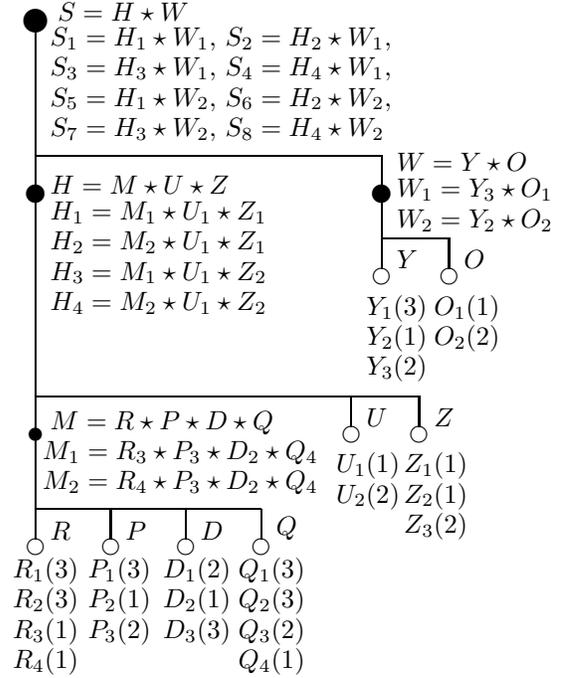
\put(03,0){\makebox(0,0)[bl]{Fig. 3. Structure of wireless sensor
element}}

\put(1,17){\makebox(0,8)[bl]{\(R_{1}(3)\)}}
\put(1,13){\makebox(0,8)[bl]{\(R_{2}(3)\)}}

\put(1,9){\makebox(0,8)[bl]{\(R_{3}(1)\)}}
\put(1,5){\makebox(0,8)[bl]{\(R_{4}(1)\)}}

\put(11,17){\makebox(0,8)[bl]{\(P_{1}(3)\)}}
\put(11,13){\makebox(0,8)[bl]{\(P_{2}(1)\)}}
\put(11,9){\makebox(0,8)[bl]{\(P_{3}(2)\)}}

\put(21,17){\makebox(0,8)[bl]{\(D_{1}(2)\)}}
\put(21,13){\makebox(0,8)[bl]{\(D_{2}(1)\)}}
\put(21,9){\makebox(0,8)[bl]{\(D_{3}(3)\)}}

\put(31,17){\makebox(0,8)[bl]{\(Q_{1}(3)\)}}

\put(31,13){\makebox(0,8)[bl]{\(Q_{2}(3)\)}}
\put(31,9){\makebox(0,8)[bl]{\(Q_{3}(2)\)}}
\put(31,5){\makebox(0,8)[bl]{\(Q_{4}(1)\)}}


\put(4,27){\line(1,0){30}}

\put(04,27){\line(0,-1){04}} \put(14,27){\line(0,-1){04}}
\put(24,27){\line(0,-1){04}} \put(34,27){\line(0,-1){04}}


\put(04,22){\circle{2}} \put(14,22){\circle{2}}
\put(24,22){\circle{2}} \put(34,22){\circle{2}}

\put(06,23){\makebox(0,8)[bl]{\(R\) }}
\put(16,23){\makebox(0,8)[bl]{\(P\) }}
\put(26,23){\makebox(0,8)[bl]{\(D\) }}
\put(36,23){\makebox(0,8)[bl]{\(Q\) }}


\put(04,27){\line(0,1){20}}

\put(04,37){\circle*{1.7}}

\put(06,37){\makebox(0,8)[bl]{\(M = R \star P \star D \star Q\)}}

\put(05,33){\makebox(0,8)[bl]{\(M_{1} = R_{3} \star P_{3} \star
D_{2} \star Q_{4}\) }}

\put(05,29){\makebox(0,8)[bl]{\(M_{2} = R_{4} \star P_{3} \star
D_{2} \star Q_{4}\) }}


\put(44,31){\makebox(0,8)[bl]{\(U_{1}(1)\)}}
\put(44,27){\makebox(0,8)[bl]{\(U_{2}(2)\)}}

\put(53,31){\makebox(0,8)[bl]{\(Z_{1}(1)\)}}
\put(53,27){\makebox(0,8)[bl]{\(Z_{2}(1)\)}}
\put(53,23){\makebox(0,8)[bl]{\(Z_{3}(2)\)}}


\put(48,38){\makebox(0,8)[bl]{\(U\) }}
\put(57,38){\makebox(0,8)[bl]{\(Z\) }}

\put(46,42){\line(0,-1){04}} \put(55,42){\line(0,-1){04}}

\put(46,37){\circle{2}} \put(55,37){\circle{2}}

\put(04,42){\line(1,0){51}}


\put(4,69){\circle*{2.5}}

\put(4,69){\line(0,-1){22}}

\put(6,69){\makebox(0,8)[bl]{\(H= M \star U \star Z\) }}

\put(6,65){\makebox(0,8)[bl]{\(H_{1} = M_{1} \star U_{1} \star
Z_{1}\) }}

\put(6,61){\makebox(0,8)[bl]{\(H_{2} = M_{2} \star U_{1} \star
Z_{1}\) }}

\put(6,57){\makebox(0,8)[bl]{\(H_{3} = M_{1} \star U_{1} \star
Z_{2}\) }}

\put(6,53){\makebox(0,8)[bl]{\(H_{4} = M_{2} \star U_{1} \star
Z_{2}\) }}


\put(48,52){\makebox(0,8)[bl]{\(Y_{1}(3)\)}}
\put(48,48){\makebox(0,8)[bl]{\(Y_{2}(1)\)}}
\put(48,44){\makebox(0,8)[bl]{\(Y_{3}(2)\)}}

\put(57,52){\makebox(0,8)[bl]{\(O_{1}(1)\)}}
\put(57,48){\makebox(0,8)[bl]{\(O_{2}(2)\)}}


\put(52,59){\makebox(0,8)[bl]{\(Y\) }}
\put(61,59){\makebox(0,8)[bl]{\(O\) }}

\put(50,63){\line(0,-1){04}} \put(59,63){\line(0,-1){04}}

\put(50,58){\circle{2}} \put(59,58){\circle{2}}

\put(50,63){\line(1,0){9}}


\put(50,59){\line(0,1){10}} \put(50,69){\circle*{2.5}}


\put(52,72){\makebox(0,8)[bl]{\(W=Y\star O\) }}
\put(52,68){\makebox(0,8)[bl]{\(W_{1}=Y_{3}\star O_{1}\) }}
\put(52,64){\makebox(0,8)[bl]{\(W_{2}=Y_{2}\star O_{2}\) }}



\put(04,69){\line(0,1){05}} \put(50,69){\line(0,1){05}}
\put(04,74){\line(1,0){46}}

\put(04,74){\line(0,1){18}}

\put(04,92){\circle*{3}}

\put(7,92){\makebox(0,8)[bl]{\(S = H \star W\) }}

\put(6,88){\makebox(0,8)[bl]{\(S_{1} = H_{1} \star W_{1}\),
 \(S_{2} = H_{2} \star W_{1}\),  }}

\put(6,84){\makebox(0,8)[bl]{\(S_{3} = H_{3} \star W_{1}\),
  \(S_{4} = H_{4} \star W_{1}\), }}

\put(6,80){\makebox(0,8)[bl]{\(S_{5} = H_{1} \star W_{2}\),
  \(S_{6} = H_{2} \star W_{2}\), }}

\put(6,76){\makebox(0,8)[bl]{\(S_{7} = H_{3} \star W_{2}\),
  \(S_{8} = H_{4} \star W_{2}\)}}

\end{picture}
\end{center}

 {\bf 0.} Alarm wireless sensor element  ~\(S = H \star W\).

  {\bf 1.} Hardware ~ \(H = M \star U \star Z\).

 {\it 1.1.} Microelectronic components ~
 \(M = R \star P \star D \star Q\).

    {\it 1.1.1.} Radio  \(R\):~
    Chipcon CC2420 Radio    \(R_{1}(3)\),
    Chipcon CC1000 Radio    \(R_{2}(4)\),
    Semtech XE1205 Radio   \(R_{3}(2)\),
    Infineon  TDA5250 Radio \(R_{4}(1)\).

    {\it 1.1.2.} Microprocessor  \(P\):~
  Atmel ATmega128 with 10-bit ADC   \(P_{1}(3)\),
  Atmel AVR AT90S2313    \(P_{2}(1)\),
   Texas Instruments MSP430F16 with 12-bit ADC/DAC    \(P_{3}(2)\).

    {\it 1.1.3.} DAC/ADC  \(D\):~
   Atmel ATmega128L embedded 10-bit ADC     \(D_{1}(2)\),
    Texas Instruments MSP430F16 embedded 12-bit ADC/DAC     \(D_{2}(1)\),
   Analog Devices 14-bit AD679   \(D_{3}(3)\).

    {\it 1.1.4.} Memory  \(Q\):~
    No external memory  \(Q_{1}(4)\),
    4 Kb EEPROM  \(Q_{2}(3)\),
    128 Kb Flash \(Q_{3}(2)\),
    1 Mb Flash \(Q_{4}(1)\).

   {\it 1.2.} Power supply  \(U\):~
   2800 mAh NiMh Battery   \(U_{1}(1)\),
   1500 mAh Li-Ion Battery \(U_{2}(2)\).

   {\it 1.3.} Sensor  \(Z\):~
   Edwards 284b-pl Heat Detector   \(Z_{1}(1)\),
   123 Security Systems Photoelectric 2-Wire Smoke  \(Z_{2}(2)\),
   Multisensing Fire Detector  \(Z_{3}(3)\).

   {\bf 2.} Software~  \(W = Y \star O\).

   {\it 2.1.} Sensor software  \(Y\):~
   Zigbee/802.15.4 \& Application \(Y_{1}(3)\),
   TinyOS BMAC \& Application   \(Y_{2}(1)\),
   Ad-Hoc software \& Application~ \(Y_{3}(2)\).

   {\it 2.2.} OS \(O\):~
   No OS, Simple run-time environment   \(O_{1}(1)\),
   TinyOS~ \(O_{2}(2)\).

 \begin{center}
\begin{picture}(73,114)
\put(04.5,110){\makebox(0,0)[bl]{Table 1. Estimates of DAs upon
criteria}}

\put(00,0){\line(1,0){73}} \put(00,98){\line(1,0){73}}
\put(00,108){\line(1,0){73}}

\put(00,0){\line(0,1){108}} \put(09,0){\line(0,1){108}}
\put(73,0){\line(0,1){108}}

\put(01,94){\makebox(0,0)[bl]{\(R_{1}\)}}
\put(01,90){\makebox(0,0)[bl]{\(R_{2}\)}}
\put(01,86){\makebox(0,0)[bl]{\(R_{3}\)}}
\put(01,82){\makebox(0,0)[bl]{\(R_{4}\)}}

\put(01,78){\makebox(0,0)[bl]{\(P_{1}\)}}
\put(01,74){\makebox(0,0)[bl]{\(P_{2}\)}}
\put(01,70){\makebox(0,0)[bl]{\(P_{3}\)}}

\put(01,66){\makebox(0,0)[bl]{\(D_{1}\)}}
\put(01,62){\makebox(0,0)[bl]{\(D_{2}\)}}
\put(01,58){\makebox(0,0)[bl]{\(D_{3}\)}}

\put(01,54){\makebox(0,0)[bl]{\(Q_{1}\)}}
\put(01,50){\makebox(0,0)[bl]{\(Q_{2}\)}}
\put(01,46){\makebox(0,0)[bl]{\(Q_{3}\)}}

\put(01,42){\makebox(0,0)[bl]{\(Q_{4}\)}}

\put(01,38){\makebox(0,0)[bl]{\(U_{1}\)}}
\put(01,34){\makebox(0,0)[bl]{\(U_{2}\)}}

\put(01,30){\makebox(0,0)[bl]{\(Z_{1}\)}}
\put(01,26){\makebox(0,0)[bl]{\(Z_{2}\)}}
\put(01,22){\makebox(0,0)[bl]{\(Z_{3}\)}}

\put(01,18){\makebox(0,0)[bl]{\(Y_{1}\)}}
\put(01,14){\makebox(0,0)[bl]{\(Y_{2}\)}}
\put(01,10){\makebox(0,0)[bl]{\(Y_{3}\)}}

\put(01,06){\makebox(0,0)[bl]{\(O_{1}\)}}
\put(01,02){\makebox(0,0)[bl]{\(O_{2}\)}}

\put(16,98){\line(0,1){10}} \put(24,98){\line(0,1){10}}

\put(31,98){\line(0,1){10}} \put(38,98){\line(0,1){10}}
\put(44,98){\line(0,1){10}} \put(58,98){\line(0,1){10}}
\put(64,98){\line(0,1){10}}

\put(11,104){\makebox(0,0)[bl]{\(C_{1}\)}}

\put(18,104){\makebox(0,0)[bl]{\(C_{2}\)}}
\put(25,104){\makebox(0,0)[bl]{\(C_{3}\)}}
\put(32,104){\makebox(0,0)[bl]{\(C_{4}\)}}
\put(39,104){\makebox(0,0)[bl]{\(C_{5}\)}}
\put(49,104){\makebox(0,0)[bl]{\(C_{6}\)}}
\put(59,104){\makebox(0,0)[bl]{\(C_{7}\)}}

\put(64.7,104){\makebox(0,0)[bl]{Prio-}}
\put(64.7,100){\makebox(0,0)[bl]{rity}}


\put(11,94){\makebox(0,0)[bl]{\(13\)}}
\put(18,94){\makebox(0,0)[bl]{\(80\)}}

\put(25.5,94){\makebox(0,0)[bl]{\(25\)}}

\put(32,94){\makebox(0,0)[bl]{\(250\)}}




\put(67,94){\makebox(0,0)[bl]{\(3\)}}


\put(11,90){\makebox(0,0)[bl]{\(11\)}}
\put(17,90){\makebox(0,0)[bl]{\(160\)}}

\put(25.5,90){\makebox(0,0)[bl]{\(29\)}}

\put(33,90){\makebox(0,0)[bl]{\(76\)}}

\put(40,90){\makebox(0,0)[bl]{\(\)}}



\put(67,90){\makebox(0,0)[bl]{\(3\)}}


\put(12,86){\makebox(0,0)[bl]{\(6\)}}
\put(17,86){\makebox(0,0)[bl]{\(600\)}}
\put(25.5,86){\makebox(0,0)[bl]{\(25\)}}

\put(33,86){\makebox(0,0)[bl]{\(76\)}}

\put(40,86){\makebox(0,0)[bl]{\(\)}}

\put(67,86){\makebox(0,0)[bl]{\(1\)}}


\put(12,82){\makebox(0,0)[bl]{\(8\)}}
\put(17,82){\makebox(0,0)[bl]{\(200\)}}
\put(25.5,82){\makebox(0,0)[bl]{\(17\)}}

\put(33,82){\makebox(0,0)[bl]{\(64\)}}


\put(67,82){\makebox(0,0)[bl]{\(1\)}}


\put(12,78){\makebox(0,0)[bl]{\(8\)}}
\put(13,78){\makebox(0,0)[bl]{\(\)}}
\put(26.5,78){\makebox(0,0)[bl]{\(8\)}}


\put(33,78){\makebox(0,0)[bl]{\(16\)}}

\put(67,78){\makebox(0,0)[bl]{\(3\)}}


\put(10.5,74){\makebox(0,0)[bl]{\(2.5\)}}
\put(13,74){\makebox(0,0)[bl]{\(\)}}
\put(26.5,74){\makebox(0,0)[bl]{\(5\)}}

\put(33,74){\makebox(0,0)[bl]{\(10\)}}


\put(67,74){\makebox(0,0)[bl]{\(1\)}}


\put(11,70){\makebox(0,0)[bl]{\(11\)}}
\put(13,70){\makebox(0,0)[bl]{\(\)}}
\put(26.5,70){\makebox(0,0)[bl]{\(2\)}}

\put(33,70){\makebox(0,0)[bl]{\(12\)}}


\put(67,70){\makebox(0,0)[bl]{\(2\)}}


\put(12,66){\makebox(0,0)[bl]{\(0\)}}
\put(26.5,66){\makebox(0,0)[bl]{\(2\)}}

\put(32,66){\makebox(0,0)[bl]{\(150\)}}


\put(39.5,66){\makebox(0,0)[bl]{\(10\)}}

\put(67,66){\makebox(0,0)[bl]{\(2\)}}


\put(12,62){\makebox(0,0)[bl]{\(0\)}}
\put(26.5,62){\makebox(0,0)[bl]{\(1\)}}

\put(32,62){\makebox(0,0)[bl]{\(200\)}}

\put(39.5,62){\makebox(0,0)[bl]{\(12\)}}

\put(67,62){\makebox(0,0)[bl]{\(1\)}}


\put(12,58){\makebox(0,0)[bl]{\(4\)}}
\put(26.5,58){\makebox(0,0)[bl]{\(4\)}}

\put(32,58){\makebox(0,0)[bl]{\(250\)}}

\put(39.5,58){\makebox(0,0)[bl]{\(14\)}}

\put(67,58){\makebox(0,0)[bl]{\(3\)}}


\put(12,54){\makebox(0,0)[bl]{\(0\)}}
\put(26.5,54){\makebox(0,0)[bl]{\(0\)}}

\put(33.5,54){\makebox(0,0)[bl]{\(0\)}}


\put(50,54){\makebox(0,0)[bl]{\(0\)}}


\put(67,54){\makebox(0,0)[bl]{\(3\)}}


\put(12,50){\makebox(0,0)[bl]{\(1\)}}
\put(26.5,50){\makebox(0,0)[bl]{\(2\)}}


\put(33.5,50){\makebox(0,0)[bl]{\(3\)}}

\put(47,50){\makebox(0,0)[bl]{\(1024\)}}


\put(67,50){\makebox(0,0)[bl]{\(3\)}}


\put(12,46){\makebox(0,0)[bl]{\(3\)}}
\put(26.5,46){\makebox(0,0)[bl]{\(3\)}}

\put(33.5,46){\makebox(0,0)[bl]{\(2\)}}

\put(45.5,46){\makebox(0,0)[bl]{\(131072\)}}


\put(67,46){\makebox(0,0)[bl]{\(2\)}}


\put(12,42){\makebox(0,0)[bl]{\(3\)}}
\put(26.5,42){\makebox(0,0)[bl]{\(3\)}}

\put(33.5,42){\makebox(0,0)[bl]{\(2\)}}

\put(45,42){\makebox(0,0)[bl]{\(1048576\)}}


\put(67,42){\makebox(0,0)[bl]{\(1\)}}


\put(12,38){\makebox(0,0)[bl]{\(3\)}}


\put(47.5,38){\makebox(0,0)[bl]{\(2800\)}}


\put(67,38){\makebox(0,0)[bl]{\(1\)}}


\put(11,34){\makebox(0,0)[bl]{\(10\)}}

\put(47.5,34){\makebox(0,0)[bl]{\(1500\)}}

\put(67,34){\makebox(0,0)[bl]{\(2\)}}


\put(11,30){\makebox(0,0)[bl]{\(10\)}}

\put(40,30){\makebox(0,0)[bl]{\(2\)}}

\put(67,30){\makebox(0,0)[bl]{\(1\)}}


\put(11,26){\makebox(0,0)[bl]{\(25\)}}

\put(40,26){\makebox(0,0)[bl]{\(5\)}}

\put(67,26){\makebox(0,0)[bl]{\(1\)}}


\put(11,22){\makebox(0,0)[bl]{\(50\)}}

\put(39.5,22){\makebox(0,0)[bl]{\(16\)}}

\put(67,22){\makebox(0,0)[bl]{\(3\)}}


\put(10,18){\makebox(0,0)[bl]{\(100\)}}

\put(47,18){\makebox(0,0)[bl]{\(15000\)}}
\put(60,18){\makebox(0,0)[bl]{\(5\)}}

\put(67,18){\makebox(0,0)[bl]{\(3\)}}


\put(11,14){\makebox(0,0)[bl]{\(50\)}}

\put(47.5,14){\makebox(0,0)[bl]{\(6000\)}}
\put(60,14){\makebox(0,0)[bl]{\(6\)}}

\put(67,14){\makebox(0,0)[bl]{\(1\)}}


\put(10,10){\makebox(0,0)[bl]{\(100\)}}

\put(32,10){\makebox(0,0)[bl]{\(\)}}
\put(47.5,10){\makebox(0,0)[bl]{\(4000\)}}
\put(59,10){\makebox(0,0)[bl]{\(11\)}}

\put(67,10){\makebox(0,0)[bl]{\(2\)}}


\put(12,06){\makebox(0,0)[bl]{\(0\)}}

\put(47.5,06){\makebox(0,0)[bl]{\(2000\)}}
\put(60,06){\makebox(0,0)[bl]{\(4\)}}

\put(67,06){\makebox(0,0)[bl]{\(1\)}}


\put(12,2){\makebox(0,0)[bl]{\(0\)}}

\put(47.5,2){\makebox(0,0)[bl]{\(4500\)}}
\put(60,2){\makebox(0,0)[bl]{\(0\)}}

\put(67,2){\makebox(0,0)[bl]{\(2\)}}


\end{picture}
\end{center}


 The following generalized set of criteria for DAs is used
(criteria weights are shown in parentheses, symbol \(-\)
corresponds to the case when minimum value is the best one):
 cost \(C_{1}\) (-100),
 radius \(C_{2}\) (1),
 power consumption  \(C_{3}\)  (-80),
  speed/frequency  \(C_{4}\)  (1),
  fidelity  \(C_{5}\)  (10),
  capacity(memory)  \(C_{6}\)  (0.5), and
  development duration  \(C_{7}\) (1000).
 Estimates of DAs upon the criteria
 are presented in Table 1 (expert judgment).

  The resultant priorities of DAs
   are pointed out in Fig. 3
  (priorities are shown in parentheses).
 and in Table 1
 (a modification
 of  outranking  technique ELECTRE was used \cite{roy96}).

  Table 2 and Table 3 contain estimates of  compatibility between DAs.
 Mainly, estimates are illustrative ones.
 For components of \(M\), \(U\)  and \(S\) equal compatibility estimates
 (between corresponding local DAs) are considered.

\begin{center}
\begin{picture}(67,54)
\put(16,50){\makebox(0,0)[bl]{Table 2. Compatibility }}

\put(00,0){\line(1,0){67}} \put(00,42){\line(1,0){67}}
\put(00,48){\line(1,0){67}}

\put(00,0){\line(0,1){48}} \put(07,0){\line(0,1){48}}
\put(67,0){\line(0,1){48}}

\put(01,38){\makebox(0,0)[bl]{\(R_{1}\)}}
\put(01,34){\makebox(0,0)[bl]{\(R_{2}\)}}
\put(01,30){\makebox(0,0)[bl]{\(R_{3}\)}}
\put(01,26){\makebox(0,0)[bl]{\(R_{4}\)}}

\put(01,22){\makebox(0,0)[bl]{\(P_{1}\)}}
\put(01,18){\makebox(0,0)[bl]{\(P_{2}\)}}
\put(01,14){\makebox(0,0)[bl]{\(P_{3}\)}}

\put(01,10){\makebox(0,0)[bl]{\(D_{1}\)}}
\put(01,06){\makebox(0,0)[bl]{\(D_{2}\)}}
\put(01,02){\makebox(0,0)[bl]{\(D_{3}\)}}

\put(13,42){\line(0,1){6}} \put(19,42){\line(0,1){6}}
\put(25,42){\line(0,1){6}} \put(31,42){\line(0,1){6}}
\put(37,42){\line(0,1){6}} \put(43,42){\line(0,1){6}}
\put(49,42){\line(0,1){6}} \put(55,42){\line(0,1){6}}
\put(61,42){\line(0,1){6}}

\put(08,44){\makebox(0,0)[bl]{\(P_{1}\)}}
\put(14,44){\makebox(0,0)[bl]{\(P_{2}\)}}
\put(20,44){\makebox(0,0)[bl]{\(P_{3}\)}}
\put(26,44){\makebox(0,0)[bl]{\(D_{1}\)}}
\put(32,44){\makebox(0,0)[bl]{\(D_{2}\)}}
\put(38,44){\makebox(0,0)[bl]{\(D_{3}\)}}

\put(44,44){\makebox(0,0)[bl]{\(Q_{1}\)}}
\put(50,44){\makebox(0,0)[bl]{\(Q_{2}\)}}
\put(56,44){\makebox(0,0)[bl]{\(Q_{3}\)}}
\put(62,44){\makebox(0,0)[bl]{\(Q_{4}\)}}

\put(09,38){\makebox(0,0)[bl]{\(3\)}}
\put(15,38){\makebox(0,0)[bl]{\(3\)}}
\put(21,38){\makebox(0,0)[bl]{\(3\)}}
\put(27,38){\makebox(0,0)[bl]{\(3\)}}
\put(33,38){\makebox(0,0)[bl]{\(3\)}}
\put(39,38){\makebox(0,0)[bl]{\(3\)}}

\put(45,38){\makebox(0,0)[bl]{\(3\)}}
\put(51,38){\makebox(0,0)[bl]{\(3\)}}
\put(57,38){\makebox(0,0)[bl]{\(3\)}}
\put(63,38){\makebox(0,0)[bl]{\(3\)}}

\put(09,34){\makebox(0,0)[bl]{\(3\)}}
\put(15,34){\makebox(0,0)[bl]{\(3\)}}
\put(21,34){\makebox(0,0)[bl]{\(3\)}}
\put(27,34){\makebox(0,0)[bl]{\(3\)}}
\put(33,34){\makebox(0,0)[bl]{\(3\)}}
\put(39,34){\makebox(0,0)[bl]{\(3\)}}

\put(45,34){\makebox(0,0)[bl]{\(3\)}}
\put(51,34){\makebox(0,0)[bl]{\(3\)}}
\put(57,34){\makebox(0,0)[bl]{\(3\)}}
\put(63,34){\makebox(0,0)[bl]{\(3\)}}

\put(09,30){\makebox(0,0)[bl]{\(3\)}}
\put(15,30){\makebox(0,0)[bl]{\(3\)}}
\put(21,30){\makebox(0,0)[bl]{\(3\)}}
\put(27,30){\makebox(0,0)[bl]{\(3\)}}
\put(33,30){\makebox(0,0)[bl]{\(3\)}}
\put(39,30){\makebox(0,0)[bl]{\(3\)}}

\put(45,30){\makebox(0,0)[bl]{\(3\)}}
\put(51,30){\makebox(0,0)[bl]{\(3\)}}
\put(57,30){\makebox(0,0)[bl]{\(3\)}}
\put(63,30){\makebox(0,0)[bl]{\(3\)}}

\put(09,26){\makebox(0,0)[bl]{\(3\)}}
\put(15,26){\makebox(0,0)[bl]{\(3\)}}
\put(21,26){\makebox(0,0)[bl]{\(3\)}}
\put(27,26){\makebox(0,0)[bl]{\(3\)}}
\put(33,26){\makebox(0,0)[bl]{\(3\)}}
\put(39,26){\makebox(0,0)[bl]{\(3\)}}

\put(45,26){\makebox(0,0)[bl]{\(3\)}}
\put(51,26){\makebox(0,0)[bl]{\(3\)}}
\put(57,26){\makebox(0,0)[bl]{\(3\)}}
\put(63,26){\makebox(0,0)[bl]{\(3\)}}

\put(09,22){\makebox(0,0)[bl]{\(\)}}
\put(15,22){\makebox(0,0)[bl]{\(\)}}
\put(21,22){\makebox(0,0)[bl]{\(\)}}
\put(27,22){\makebox(0,0)[bl]{\(3\)}}
\put(33,22){\makebox(0,0)[bl]{\(0\)}}
\put(39,22){\makebox(0,0)[bl]{\(1\)}}

\put(45,22){\makebox(0,0)[bl]{\(3\)}}
\put(51,22){\makebox(0,0)[bl]{\(3\)}}
\put(57,22){\makebox(0,0)[bl]{\(3\)}}
\put(63,22){\makebox(0,0)[bl]{\(3\)}}

\put(09,18){\makebox(0,0)[bl]{\(\)}}
\put(15,18){\makebox(0,0)[bl]{\(\)}}
\put(21,18){\makebox(0,0)[bl]{\(\)}}
\put(27,18){\makebox(0,0)[bl]{\(0\)}}
\put(33,18){\makebox(0,0)[bl]{\(0\)}}
\put(39,18){\makebox(0,0)[bl]{\(1\)}}

\put(45,18){\makebox(0,0)[bl]{\(3\)}}
\put(51,18){\makebox(0,0)[bl]{\(3\)}}
\put(57,18){\makebox(0,0)[bl]{\(3\)}}
\put(63,18){\makebox(0,0)[bl]{\(3\)}}

\put(09,14){\makebox(0,0)[bl]{\(\)}}
\put(15,14){\makebox(0,0)[bl]{\(\)}}
\put(21,14){\makebox(0,0)[bl]{\(\)}}
\put(27,14){\makebox(0,0)[bl]{\(0\)}}
\put(33,14){\makebox(0,0)[bl]{\(3\)}}
\put(39,14){\makebox(0,0)[bl]{\(1\)}}

\put(45,14){\makebox(0,0)[bl]{\(3\)}}
\put(51,14){\makebox(0,0)[bl]{\(3\)}}
\put(57,14){\makebox(0,0)[bl]{\(3\)}}
\put(63,14){\makebox(0,0)[bl]{\(3\)}}



\put(45,10){\makebox(0,0)[bl]{\(3\)}}
\put(51,10){\makebox(0,0)[bl]{\(3\)}}
\put(57,10){\makebox(0,0)[bl]{\(3\)}}
\put(63,10){\makebox(0,0)[bl]{\(3\)}}


\put(45,6){\makebox(0,0)[bl]{\(3\)}}
\put(51,6){\makebox(0,0)[bl]{\(3\)}}
\put(57,6){\makebox(0,0)[bl]{\(3\)}}
\put(63,6){\makebox(0,0)[bl]{\(3\)}}


\put(45,2){\makebox(0,0)[bl]{\(3\)}}
\put(51,2){\makebox(0,0)[bl]{\(3\)}}
\put(57,2){\makebox(0,0)[bl]{\(3\)}}
\put(63,2){\makebox(0,0)[bl]{\(3\)}}

\end{picture}
\end{center}

\begin{center}
\begin{picture}(40,26)

\put(02,22){\makebox(0,0)[bl]{Table 3. Compatibility}}

\put(10,0){\line(1,0){19}} \put(10,14){\line(1,0){19}}
\put(10,20){\line(1,0){19}}

\put(10,0){\line(0,1){20}} \put(17,0){\line(0,1){20}}
\put(29,0){\line(0,1){20}}

\put(11,10){\makebox(0,0)[bl]{\(Y_{1}\)}}
\put(11,06){\makebox(0,0)[bl]{\(Y_{2}\)}}
\put(11,02){\makebox(0,0)[bl]{\(Y_{3}\)}}

\put(23,14){\line(0,1){6}}

\put(18,16){\makebox(0,0)[bl]{\(O_{1}\)}}
\put(24,16){\makebox(0,0)[bl]{\(O_{2}\)}}

\put(19,10){\makebox(0,0)[bl]{\(1\)}}
\put(25,10){\makebox(0,0)[bl]{\(2\)}}

\put(19,6){\makebox(0,0)[bl]{\(0\)}}
\put(25,6){\makebox(0,0)[bl]{\(3\)}}

\put(19,2){\makebox(0,0)[bl]{\(3\)}}
\put(25,2){\makebox(0,0)[bl]{\(2\)}}

\end{picture}
\end{center}

\section{Combinatorial Synthesis}


 Second,
 Hierarchical Morphological Multicriteria Design (HMMD)
 based on morphological clique problem is considered
 (e.g., \cite{lev98},
 \cite{lev06},
 \cite{lev09},
 \cite{lev12}).
 HMMD generalizes morphological analysis that was suggested by
 F. Zwicky
 \cite{zwi69}.
 Development stages of morphological analysis based design approaches
  are presented in  \cite{lev12}.

 A examined composite
 (modular, decomposable, composable) system consists
 of components and their interconnection or compatibility (IC).
 Basic assumptions of HMMD are the following:
 ~(a) a tree-like structure of the system;
 ~(b) a composite estimate for system quality
     that integrates components (subsystems, parts) qualities and
    qualities of IC (compatibility) across subsystems;
 ~(c) monotonic criteria for the system and its components;
 ~(d) quality estimates of system components and IC are evaluated by
 coordinated ordinal scales.
 The designations are:
  ~(1) design alternatives (DAs) for
  nodes of the model;
  ~(2) priorities of DAs (\(r=\overline{1,k}\);
      \(1\) corresponds to the best level of quality);
  ~(3) an ordinal compatibility estimate for each pair of DAs
  (\(w=\overline{0,l}\); \(l\) corresponds to the best level of quality).
 Generally, the basic phases of HMMD are:

  {\bf 1.} Design of the tree-like system model.

  {\bf 2.} Generation of DAs for leaf nodes of the model.

  {\bf 3.} Hierarchical selection and composing of DAs into composite
    DAs for the corresponding higher level of the system
    hierarchy.

  {\bf 4.} Analysis and improvement of composite DAs (solution(s)).

 Let \(S\) be a system consisting of \(m\) parts (components):
 \(P(1),...,P(i),...,P(m)\).
 A set of design alternatives (DAs)
 is generated for each system part above.
 The problem is:

 {\it Find composite design alternative}
 ~ \(S=S(1)\star ...\star S(i)\star ...\star S(m)\)~
 ({\it one representative design alternative} \(S(i)\)
 {\it for each system component/part} ~\(P(i)\), \(i=\overline{1,m}\))
 {\it with non-zero}~ IC {\it estimates between
 the representative
 design
 alternatives.}

 A discrete space of the integrated system excellence is based
 on the
 following vector:
 ~~\(N(S)=(w(S);n(S))\),
 ~where \(w(S)\) is the minimum of pairwise compatibility
 between DAs which correspond to different system components
 (i.e.,
 \(~\forall ~P_{j_{1}}\) and \( P_{j_{2}}\),
 \(1 \leq j_{1} \neq j_{2} \leq m\))
 in \(S\),
 ~\(n(S)=(n_{1},...,n_{r},...n_{k})\),
 ~where \(n_{r}\) is the number of DAs of the \(r\)th quality in \(S\)
 ~(\(\sum^{k}_{r=1} n_{r} = m\)).
 As a result,
 we search for composite decisions which are nondominated by \(N(S)\)
 (i.e., Pareto-efficient solutions).
%
%
%
%
%
 Fig. 4 depicts the lattice of system quality
 (by elements;  \(m=3\),\(k=3\)).

\begin{center}
\begin{picture}(60,81)
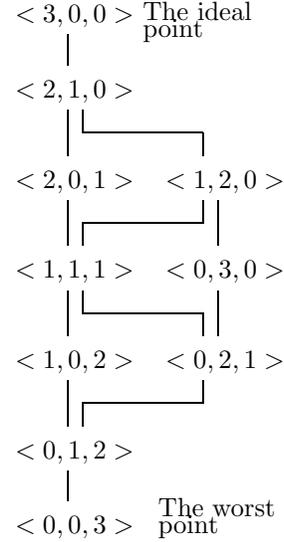
\put(0,0){\makebox(0,0)[bl] {Fig. 4. Lattice of quality (by
elements)}}

\put(27,74){\makebox(0,0)[bl]{The ideal}}
\put(27,71){\makebox(0,0)[bl]{point}}

\put(10,73){\makebox(0,0)[bl]{\(<3,0,0>\) }}

\put(17,68){\line(0,1){4}}
\put(10,63){\makebox(0,0)[bl]{\(<2,1,0>\)}}

\put(17,56){\line(0,1){6}}
\put(10,51){\makebox(0,0)[bl]{\(<2,0,1>\) }}

\put(17,44){\line(0,1){6}}
\put(10,39){\makebox(0,0)[bl]{\(<1,1,1>\) }}

\put(17,32){\line(0,1){6}}
\put(10,27){\makebox(0,0)[bl]{\(<1,0,2>\) }}


\put(17,20){\line(0,1){6}}
\put(10,15){\makebox(0,0)[bl]{\(<0,1,2>\) }}

\put(17,10){\line(0,1){4}}

\put(10,05){\makebox(0,0)[bl]{\(<0,0,3>\) }}
\put(29,08){\makebox(0,0)[bl]{The worst}}
\put(29,05){\makebox(0,0)[bl]{point}}

\put(19,59){\line(0,1){3}} \put(35,59){\line(-1,0){16}}
\put(35,56){\line(0,1){3}}

\put(30,51){\makebox(0,0)[bl]{\(<1,2,0>\) }}

\put(35,50){\line(0,-1){3}} \put(35,47){\line(-1,0){16}}
\put(19,47){\line(0,-1){3}}
\put(37,44){\line(0,1){6}}
\put(30,39){\makebox(0,0)[bl]{\(<0,3,0>\) }}

\put(19,35){\line(0,1){3}} \put(35,35){\line(-1,0){16}}
\put(35,32){\line(0,1){3}}

\put(37,32){\line(0,1){6}}
\put(30,27){\makebox(0,0)[bl]{\(<0,2,1>\) }}

\put(35,26){\line(0,-1){3}} \put(35,23){\line(-1,0){16}}
\put(19,23){\line(0,-1){3}}

\end{picture}
\end{center}

 Now, let us consider combinatorial synthesis for
 the subsystems of wireless sensor.
 The obtained
 Pareto-efficient
 composite DAs for subsystems are the following:

 (a) \( W_{1}= Y_{3} \star O_{1}\), ~\(N(W_{1})=(3;1,1,0)\);

 (b) \( W_{2}= Y_{2} \star O_{2}\), ~\(N(W_{2})=(3;1,1,0)\);

 (c) \( M_{1} = R_{3} \star P_{3} \star D_{2} \star Q_{4}\),
  ~\(N(M_{1})=(3;3,1,0)\).

 (d) \( M_{2} = R_{4} \star P_{3} \star D_{2} \star Q_{4}\),
 ~\(N(M_{1})=(3;3,1,0)\).

 Fig. 5 and Fig. 6 illustrate
 solutions  for \(M_{1}\) and \(M_{2}\).

%
%
\begin{center}
\begin{picture}(50,58)
\put(0,0){\makebox(0,0)[bl] {Fig. 5. Concentric presentation}}


\put(0,30){\line(1,0){24}} \put(0,36){\line(1,0){24}}

\put(0,30){\line(0,1){6}}

\put(1,32){\makebox(0,0)[bl]{\(R_{1}\)}}


\put(6,32){\makebox(0,0)[bl]{\(R_{2}\)}}

\put(11,30){\line(0,1){6}} \put(13,30){\line(0,1){6}}

\put(14,32){\makebox(0,0)[bl]{\(R_{3}\)}}


\put(19,32){\makebox(0,0)[bl]{\(R_{4}\)}}

\put(24,30){\line(0,1){6}}


\put(30,30){\line(1,0){18}} \put(30,36){\line(1,0){18}}

\put(30,30){\line(0,1){6}}
\put(31,32){\makebox(0,0)[bl]{\(D_{2}\)}}
\put(36,30){\line(0,1){6}}
\put(37,32){\makebox(0,0)[bl]{\(D_{1}\)}}
\put(42,30){\line(0,1){6}}
\put(43,32){\makebox(0,0)[bl]{\(D_{3}\)}}

\put(48,30){\line(0,1){6}}



\put(24,39){\line(0,1){16}} \put(30,39){\line(0,1){16}}

\put(24,39){\line(1,0){6}}
\put(25,40.5){\makebox(0,0)[bl]{\(P_{2}\)}}
\put(24,44.5){\line(1,0){6}}
\put(25,46){\makebox(0,0)[bl]{\(P_{3}\)}}
\put(24,50){\line(1,0){6}}
\put(25,51){\makebox(0,0)[bl]{\(P_{1}\)}}
\put(24,55){\line(1,0){6}}




\put(24,06){\line(0,1){21}} \put(30,06){\line(0,1){21}}

\put(24,06){\line(1,0){6}}
\put(25,7.5){\makebox(0,0)[bl]{\(Q_{2}\)}}
\put(25,12){\makebox(0,0)[bl]{\(Q_{1}\)}}
\put(24,17){\line(1,0){6}}
\put(25,18){\makebox(0,0)[bl]{\(Q_{3}\)}}
\put(24,22){\line(1,0){6}}
\put(25,23){\makebox(0,0)[bl]{\(Q_{4}\)}}
\put(24,27){\line(1,0){6}}



\put(20,36){\line(0,1){10}} \put(20,46){\line(1,0){4}}



\put(16,36){\line(0,1){12}} \put(16,48){\line(1,0){8}}



\put(20,30){\line(0,-1){04}} \put(20,26){\line(1,0){4}}



\put(16,30){\line(0,-1){6}} \put(16,24){\line(1,0){8}}



\put(33,36){\line(0,1){10}} \put(33,46){\line(-1,0){3}}



\put(33,30){\line(0,-1){04}} \put(33,26){\line(-1,0){3}}



\put(30,24){\line(1,0){20}} \put(50,24){\line(0,1){24}}
\put(30,48){\line(1,0){20}}



\put(24,33){\line(1,0){06}}



\put(14,36){\line(0,1){21}} \put(14,57){\line(1,0){21}}
\put(35,36){\line(0,1){21}}






\end{picture}
\end{center}

\begin{center}
\begin{picture}(52,65)
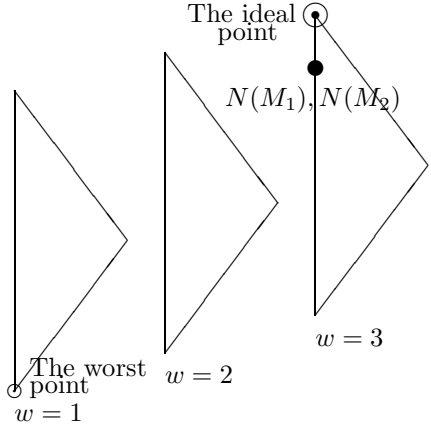
\put(02,0){\makebox(0,0)[bl]{Fig. 6. Space of system quality}}

\put(0,010){\line(0,1){40}} \put(0,010){\line(3,4){15}}
\put(0,050){\line(3,-4){15}}

\put(20,015){\line(0,1){40}} \put(20,015){\line(3,4){15}}
\put(20,055){\line(3,-4){15}}

\put(40,020){\line(0,1){40}} \put(40,020){\line(3,4){15}}
\put(40,060){\line(3,-4){15}}



\put(40,53){\circle*{2}}

\put(28,47){\makebox(0,0)[bl]{\(N(M_{1}), N(M_{2})\)}}

\put(40,60){\circle*{1}} \put(40,60){\circle{3}}


\put(23,59){\makebox(0,0)[bl]{The ideal}}
\put(27,56){\makebox(0,0)[bl]{point}}


\put(0,6){\makebox(0,0)[bl]{\(w=1\)}}
\put(20,11){\makebox(0,0)[bl]{\(w=2\)}}
\put(40,16){\makebox(0,0)[bl]{\(w=3\)}}

\put(02,12){\makebox(0,0)[bl]{The worst}}
\put(02,09){\makebox(0,0)[bl]{point}}

\put(0,10){\circle*{0.5}} \put(0,10){\circle{1.6}}

\end{picture}
\end{center}

 Further, the solutions   for \(H\) are:

  \(H_{1} = M_{1} \star U_{1} \star Z_{1} =
  (R_{3} \star P_{3} \star D_{2} \star Q_{4}) \star
  (U_{1} \star Z_{1})\);

  \(H_{2} = M_{2} \star U_{1} \star Z_{2} =
  (R_{3} \star P_{3} \star D_{2} \star Q_{4}) \star
  (U_{1} \star Z_{2})\);

 \(H_{3} = M_{1} \star U_{1} \star Z_{1} =
  (R_{4} \star P_{3} \star D_{2} \star Q_{4}) \star
  (U_{1} \star Z_{1})\);

  \(H_{4} = M_{1} \star U_{1} \star Z_{2} =
  (R_{4} \star P_{3} \star D_{2} \star Q_{4}) \star
  (U_{1} \star Z_{2})\).

  Finally, eight  resultant solutions are obtained:

 \(S_{1} = H_{1} \star W_{1}  =
 (M_{1} \star U_{1} \star Z_{1}) \star
 (Y_{3} \star O_{1})=\)
 \(
 ((R_{3} \star P_{3} \star D_{2} \star Q_{4})
 \star
 (U_{1} \star Z_{1}))
 \star
 (Y_{3} \star O_{1})\);

 \(S_{2} = H_{2} \star W_{1}  =
 (M_{2} \star U_{1} \star Z_{1}) \star
 (Y_{3} \star O_{1})=\)
 \(
 ((R_{4} \star P_{3} \star D_{2} \star Q_{4})
 \star
 (U_{1} \star Z_{1}))
 \star
 (Y_{3} \star O_{1})\);

 \(S_{3} = H_{3} \star W_{1}  =
 (M_{1} \star U_{1} \star Z_{2}) \star
 (Y_{3} \star O_{1})=\)
 \(
 ((R_{3} \star P_{3} \star D_{2} \star Q_{4})
 \star
 (U_{1} \star Z_{1}))
 \star
 (Y_{3} \star O_{1})\);

 \(S_{4} = H_{4} \star W_{1}  =
 (M_{2} \star U_{1} \star Z_{2}) \star
 (Y_{3} \star O_{1})=\)
 \(
 ((R_{4} \star P_{3} \star D_{2} \star Q_{4})
 \star
 (U_{1} \star Z_{1}))
 \star
 (Y_{3} \star O_{1})\);

 \(S_{5} = H_{1} \star W_{2}  =
 (M_{1} \star U_{1} \star Z_{1}) \star
 (Y_{2} \star O_{2})=\)
 \(
 ((R_{3} \star P_{3} \star D_{2} \star Q_{4})
 \star
 (U_{1} \star Z_{1}))
 \star
 (Y_{2} \star O_{2})\);

 \(S_{6} = H_{2} \star W_{2}  =
 (M_{2} \star U_{1} \star Z_{1}) \star
 (Y_{2} \star O_{2})=\)
 \(
 ((R_{4} \star P_{3} \star D_{2} \star Q_{4})
 \star
 (U_{1} \star Z_{1}))
 \star
 (Y_{2} \star O_{2})\);

 \(S_{7} = H_{3} \star W_{2}  =
 (M_{1} \star U_{1} \star Z_{2}) \star
 (Y_{2} \star O_{2})=\)
 \(
 ((R_{3} \star P_{3} \star D_{2} \star Q_{4})
 \star
 (U_{1} \star Z_{2}))
 \star
 (Y_{2} \star O_{2})\);

 \(S_{8} = H_{4} \star W_{2}  =
 (M_{2} \star U_{1} \star Z_{2}) \star
 (Y_{2} \star O_{2})=\)
 \(
 ((R_{4} \star P_{3} \star D_{2} \star Q_{4})
 \star
 (U_{1} \star Z_{2}))
 \star
 (Y_{2} \star O_{2})\).

 Note in the example the initial combinatorial set
  includes  ~\(5184\)  (\(4 \times 3 \times 3 \times 4 \times 2 \times 3 \times 3 \times 2\))
  possible composite   solutions.


\section{Aggregation of Modular Solutions}

 Aggregation of composite systems (as modular solutions)
 can be considered as follows
 \cite{lev11agg}.
 Fig. 7 illustrates substructure and superstructure
 for three initial solutions
 \(S^{1}\), \(S^{2}\), and \(S^{3}\).

\begin{center}
\begin{picture}(62,28)
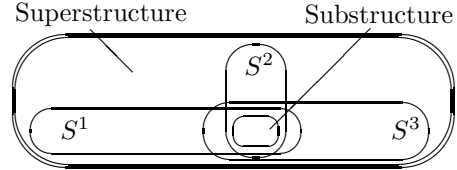
\put(00,00){\makebox(0,0)[bl]{Fig. 7. Substructure and
superstructure}}


\put(20,09){\oval(36,6)}

\put(06,8){\makebox(0,8)[bl]{\(S^{1}\)}}


\put(40,09){\oval(30,7.6)}

\put(50,8){\makebox(0,8)[bl]{\(S^{3}\)}}


\put(32,13){\oval(08,15)}

\put(30.5,16.5){\makebox(0,8)[bl]{\(S^{2}\)}}


\put(29,13){\oval(58,17)} \put(29,13){\oval(58.7,17.7)}

\put(00,23){\makebox(0,8)[bl]{Superstructure}}
 \put(10,22.5){\line(1,-1){5.6}}

\put(38.5,23.4){\makebox(0,8)[bl]{Substructure}}
 \put(47.2,22.6){\line(-1,-1){13.3}}

\put(32,09){\oval(6,4)}

\end{picture}
\end{center}

 In  \cite{lev11agg},
  basic aggregation strategies are described, for example:


 {\bf 1.} {\it Extension strategy}:
 ~{\it 1.1.} building a ``kernel'' for initial solutions
 (i.e., substructure/subsolution or an extended subsolution),
 ~{\it 1.2.} generation of a set of additional solution elements,
 ~{\it 1.3.} selection of additional elements from the generated
 set while taking into account their ``profit'' and resource requirements
 (i.e., a total ``profit'' and total resource constraint(s))
 (here knapsack-like problems are used).


  {\bf 2.} {\it Compression strategy}:
 ~{\it 2.1.} building a supersolution (as a superstructure),
 ~{\it 2.2.} generation of a set of solution elements from the
 supersolution as candidates for deletion,
 ~{\it 2.3.} selection of the  elements-candidates for deletion
 while taking into account their ``profit'' and resource requirements
 (i.e., a total profit and total resource constraint(s))
 (here knapsack-like problems with minimization of objective function are used).


 A general aggregation strategy has to be based on searching for a
 consensus/median solution
 \(S^{M}\) (``generalized'' median)
 for the initial solutions \( \overline{S} = \{S^{1},...,S^{n}\}\)
 (e.g.,  \cite{lev11agg}):
 \[ S^{M} = \arg ~ min_{X \in \overline{S}}  ~
 ( \sum_{i=1}^{n} ~ \rho (X, S^{i})  ),\]
  where
 \(\rho (X, Y)\) is a proximity (e.g., distance) between two solutions \(X\) and
 \(Y\).
  Mainly,
  searching for the median for many structures is usually NP-complete
  problem.
 In our case, product structures correspond to a combination of
 tree, set of DAs, their estimates, matrices of compatibility
 estimates. As a result, the proximity between the structures
 are more complicated and
 the
 ``generalized'' median problem is very complex.
  As a result, simplified (approximate) solving strategies is often used, for example
   \cite{lev11agg}:
  (a) searching for
  ``set median'' (i.e., one of the initial solutions is selected),
  (b) ``extension strategy'' above,
  (c) ``compression strategy'' above.

 In the example, eight obtained composite solutions are considered:
 ~\(S_{1}\),  \(S_{2}\),  \(S_{3}\),  \(S_{4}\),  \(S_{5}\),
 \(S_{6}\),  \(S_{7}\),  \(S_{8}\).
 The substructure of the eight solutions is
 presented in Fig. 8. This substructure is examined as
 system ``kernel'' for future extension.
  The superstructure is presented in Fig. 9.

\begin{center}
\begin{picture}(67,20)

\put(08,00){\makebox(0,0)[bl] {Fig. 8. Substructure
 (``kernel'')}}


\put(05,14){\circle*{1.5}} \put(13,14){\circle*{1.5}}
\put(21,14){\circle*{1.5}} \put(29,14){\circle*{1.5}}
\put(37,14){\circle*{1.5}} \put(45,14){\circle*{1.5}}
\put(53,14){\circle*{1.5}} \put(61,14){\circle*{1.5}}

\put(04,16){\makebox(0,0)[bl]{\(R\)}}
\put(12,16){\makebox(0,0)[bl]{\(P\)}}
\put(20,16){\makebox(0,0)[bl]{\(D\)}}
\put(28,15.5){\makebox(0,0)[bl]{\(Q\)}}

\put(35,16){\makebox(0,0)[bl]{\(U\)}}
\put(43,16){\makebox(0,0)[bl]{\(Z\)}}

\put(52,16){\makebox(0,0)[bl]{\(Y\)}}
\put(60,16){\makebox(0,0)[bl]{\(O\)}}


\put(05,10){\oval(07,08)} \put(13,10){\oval(07,08)}
\put(21,10){\oval(07,08)} \put(29,10){\oval(07,08)}
\put(37,10){\oval(07,08)} \put(45,10){\oval(07,08)}
\put(53,10){\oval(07,08)} \put(61,10){\oval(07,08)}


\put(10.8,09){\makebox(0,0)[bl]{\(P_{3}\)}}

\put(18.8,09){\makebox(0,0)[bl]{\(D_{2}\)}}

\put(26.8,09){\makebox(0,0)[bl]{\(Q_{4}\)}}

\put(34.8,09){\makebox(0,0)[bl]{\(U_{1}\)}}




\end{picture}
\end{center}

\begin{center}
\begin{picture}(67,22)

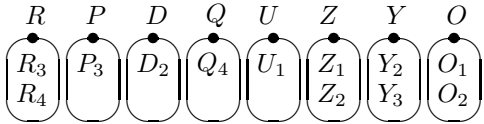
\put(06.5,00){\makebox(0,0)[bl] {Fig. 9. Superstructure of
solutions}}


\put(05,16){\circle*{1.5}} \put(13,16){\circle*{1.5}}
\put(21,16){\circle*{1.5}} \put(29,16){\circle*{1.5}}
\put(37,16){\circle*{1.5}} \put(45,16){\circle*{1.5}}
\put(53,16){\circle*{1.5}} \put(61,16){\circle*{1.5}}

\put(04,18){\makebox(0,0)[bl]{\(R\)}}
\put(12,18){\makebox(0,0)[bl]{\(P\)}}
\put(20,18){\makebox(0,0)[bl]{\(D\)}}
\put(28,17.5){\makebox(0,0)[bl]{\(Q\)}}

\put(35,18){\makebox(0,0)[bl]{\(U\)}}
\put(43,18){\makebox(0,0)[bl]{\(Z\)}}

\put(52,18){\makebox(0,0)[bl]{\(Y\)}}
\put(60,18){\makebox(0,0)[bl]{\(O\)}}


\put(05,10.5){\oval(07,11)} \put(13,10.5){\oval(07,11)}
\put(21,10.5){\oval(07,11)} \put(29,10.5){\oval(07,11)}
\put(37,10.5){\oval(07,11)} \put(45,10.5){\oval(07,11)}
\put(53,10.5){\oval(07,11)} \put(61,10.5){\oval(07,11)}

\put(02.8,11){\makebox(0,0)[bl]{\(R_{3}\)}}
\put(02.8,07){\makebox(0,0)[bl]{\(R_{4}\)}}

\put(10.8,11){\makebox(0,0)[bl]{\(P_{3}\)}}

\put(18.8,11){\makebox(0,0)[bl]{\(D_{2}\)}}

\put(26.8,11){\makebox(0,0)[bl]{\(Q_{4}\)}}

\put(34.8,11){\makebox(0,0)[bl]{\(U_{1}\)}}

\put(42.8,11){\makebox(0,0)[bl]{\(Z_{1}\)}}
\put(42.8,07){\makebox(0,0)[bl]{\(Z_{2}\)}}

\put(50.8,11){\makebox(0,0)[bl]{\(Y_{2}\)}}
\put(50.8,07){\makebox(0,0)[bl]{\(Y_{3}\)}}

\put(58.8,11){\makebox(0,0)[bl]{\(O_{1}\)}}
\put(58.8,07){\makebox(0,0)[bl]{\(O_{2}\)}}

\end{picture}
\end{center}

 The extension procedure based on multiple choice problem
   is the following.
 Table 4 contains design alternatives (DAs)
 and their estimates (ordinal scales, expert judgment).
 The design alternatives correspond to superstructure (Fig. 9).

\begin{center}
\begin{picture}(52,52)

\put(03,48){\makebox(0,0)[bl] {Table 4. Design alternatives}}

\put(0,0){\line(1,0){52}} \put(0,35){\line(1,0){52}}
\put(0,45){\line(1,0){52}}


\put(0,0){\line(0,1){45}} \put(05,0){\line(0,1){45}}
\put(18,0){\line(0,1){45}} \put(32,0){\line(0,1){45}}
\put(42,0){\line(0,1){45}} \put(52,0){\line(0,1){45}}

\put(2,41){\makebox(0,0)[bl]{\(\kappa\)}}

\put(7.5,40.5){\makebox(0,0)[bl]{DAs}}

\put(19,40.5){\makebox(0,0)[bl]{Binary}}
\put(19,37){\makebox(0,0)[bl]{variable}}

\put(33,41){\makebox(0,0)[bl]{Cost}}
\put(35,36.5){\makebox(0,0)[bl]{\(a_{ij}\)}}

\put(42.6,41){\makebox(0,0)[bl]{Profit}}
\put(45,36.5){\makebox(0,0)[bl]{\(c_{ij}\)}}




\put(1.6,30){\makebox(0,0)[bl]{\(1\)}}

\put(9,29.5){\makebox(0,0)[bl]{\(R_{3}\)}}

\put(23,30){\makebox(0,0)[bl]{\(x_{11}\)}}


\put(1.6,26){\makebox(0,0)[bl]{\(2\)}}

\put(9,25.5){\makebox(0,0)[bl]{\(R_{4}\) }}

\put(23,26){\makebox(0,0)[bl]{\(x_{12}\)}}


\put(1.6,22){\makebox(0,0)[bl]{\(3\)}}

\put(9,21.5){\makebox(0,0)[bl]{\(Z_{1} \)}}

\put(23,022){\makebox(0,0)[bl]{\(x_{21}\)}}


\put(1.6,18){\makebox(0,0)[bl]{\(4\)}}

\put(9,17.5){\makebox(0,0)[bl]{\(Z_{2} \)}}

\put(23,18){\makebox(0,0)[bl]{\(x_{22}\)}}


\put(1.6,14){\makebox(0,0)[bl]{\(5\)}}

\put(9,013.5){\makebox(0,0)[bl]{\(Y_{2}\)}}

\put(23,14){\makebox(0,0)[bl]{\(x_{31}\)}}


\put(1.6,10){\makebox(0,0)[bl]{\(6\)}}

\put(9,9.5){\makebox(0,0)[bl]{\(Y_{3}\) }}

\put(23,10){\makebox(0,0)[bl]{\(x_{32}\)}}


\put(1.6,6){\makebox(0,0)[bl]{\(7\)}}

\put(9,05.5){\makebox(0,0)[bl]{\(O_{1} \)}}

\put(23,06){\makebox(0,0)[bl]{\(x_{41}\)}}


\put(1.6,2){\makebox(0,0)[bl]{\(8\)}}

\put(9,01.5){\makebox(0,0)[bl]{\(O_{2} \)}}

\put(23,02){\makebox(0,0)[bl]{\(x_{42}\)}}



\put(36,30){\makebox(0,0)[bl]{\(2\)}}
\put(46,30){\makebox(0,0)[bl]{\(3\)}}


\put(36,26){\makebox(0,0)[bl]{\(3\)}}
\put(46,26){\makebox(0,0)[bl]{\(4\)}}


\put(36,22){\makebox(0,0)[bl]{\(4\)}}
\put(46,22){\makebox(0,0)[bl]{\(3\)}}


\put(36,18){\makebox(0,0)[bl]{\(6\)}}
\put(46,18){\makebox(0,0)[bl]{\(3\)}}


\put(36,14){\makebox(0,0)[bl]{\(7\)}}
\put(46,14){\makebox(0,0)[bl]{\(3\)}}


\put(36,10){\makebox(0,0)[bl]{\(8\)}}
\put(46,10){\makebox(0,0)[bl]{\(2\)}}


\put(36,06){\makebox(0,0)[bl]{\(1\)}}
\put(46,06){\makebox(0,0)[bl]{\(3\)}}


\put(36,02){\makebox(0,0)[bl]{\(1\)}}
\put(46,02){\makebox(0,0)[bl]{\(2\)}}

\end{picture}
\end{center}

 It is assumed design alternatives for different
 product components are compatible.
 The multiple choice problem is:
 \[\max \sum_{i=1}^{4}  \sum_{j=1}^{q_{i}}   c_{ij} x_{ij}
 ~~~s.t.~ \sum_{i=1}^{4}  \sum_{j=1}^{q_{i}}   a_{ij} x_{ij} \leq
 b,
 \]
 \[\sum_{j=1}^{q_{i}}   x_{ij} = 1 ~~  \forall i=\overline{1,4},
  ~~x_{ij} \in \{0,1\}.\]
%
%
 Clearly, \(q_{1} = 2\), \(q_{2} = 2\), \(q_{3} = 2\), \(q_{4} =
 2\).
 The  resultant aggregated solutions  are
  (a simple greedy algorithm was used;
 the algorithm is based on ordering of elements by
  \(c_{i}/a_{i}\)):

  (1) \(b^{1}=14\):~
 (\(x_{11} = 1\), \(x_{21} = 1\), \(x_{31} = 1\), \(x_{41} = 1\)),
 ~\(S^{agg}_{b^{1}} =
 R_{3}\star P_{3} \star D_{2}\star Q_{4}\star U_{1}\star Z_{1}
 \star Y_{2}\star O_{1} \);

 (2) \(b^{2}=15\):~
 (\(x_{12} = 1\), \(x_{21} = 1\), \(x_{31} = 1\), \(x_{41} = 1\)),
 ~\(S^{agg}_{b^{2}} =
 R_{4}\star P_{3} \star D_{2}\star Q_{4}\star U_{1}\star Z_{1}
 \star Y_{2}\star O_{1} \).


\section{Conclusion}

 In the article, hierarchical combinatorial approach to configuration of
 modular wireless sensor has been described.
 The solving framework is based on
 hierarchical model of the sensor element,
 morphological design method for combinatorial synthesis
 (building a special morphological clique), and
 aggregation of the obtained modular solutions
 (multiple choice problem).
 The suggested approach can be used for many modular systems.
 In the future it may be prospective  to consider
 the following research directions:
  {\it 1.} taking into account uncertainty;
%
 {\it 2.} analysis of dynamical design problems;
 {\it 3.} usage of AI solving techniques'
 and
 {\it 4.} usage of the described application and solving approach
  in engineering/CS  education.

 The draft material for the article was prepared
 within framework of faculty course ``{\it Design of systems: structural
 approach}'' at Moscow Inst. of Physics and Technology (State Univ.)
 (creator and lecturer: M.Sh. Levin, 2004...2008)
 \cite{lev11edu}.

%


\begin{thebibliography}{200}

 \bibitem {aky02}  F. Akyildiz, W. Su, Y. Sankarasubramaniam, E. Cayirci,
 Wireless sensor networks: a survey.
  Comp. Netw., 38(4) (2002) 393-422.


 \bibitem {art91}  E.I. Artamonov, V.M. Hachumov,
  Synthesis of structures for special tools of machine graphics.
 Inst. of Control Problems, Moscow, 1991 (in Russian).



 \bibitem {campbell03} M.I. Campbell, J. Cagan, K. Kotovsky,
 The A-design approach to managing automated design synthesis.
  Res. in Eng. Des.,   14(1) (2003) 12-24.

 \bibitem {ciuksys07} D. Ciuksys, A. Caplinkas,
 Reusing ontological knowledge about business processes in IS
 engineering: process configuration problem.
  Informatica, 18(4)  (2007) 585-602.

  \bibitem {cul04}  D. Culler, D. Estrin, M. Srivastava,
  Overview of sensor networks.
   IEEE Computer, 37(8) (2004) 41-49.

 \bibitem {kul11} R.V. Kulkarni, A. Forster, G.K. Venayagamoorthy,
 Computational intelligence in wireless sensor networks: A survey.
 IEEE Communications Surveys \& Tutorials,
 13(1) (2011) 68-96.

  \bibitem {lev98} M.Sh. Levin,
  Combinatorial Engineering of Decomposable Systems,
  Kluwer Academic Publishers, Dordrecht, 1998.

  \bibitem {lev06} M.Sh. Levin,
   Composite Systems Decisions,  Springer, New York, 2006.





 \bibitem {lev09} M.Sh. Levin,
 Combinatorial optimization in system configuration design,
  Autom.\&Rem. Control 70(3) (2009) 519-561.


 \bibitem {lev11edu} M.Sh. Levin,
  Course on system design (structural approach).
  Elect. Preprint, (2011) 22 pp.
   http://arxiv.org/abs/1103.3845

 \bibitem {lev11agg} M.Sh. Levin,
  Towards aggregation of composite solutions:
  strategies, models, examples.
  Elect. Preprint,
 (2011) 72 pp.
   http://arxiv.org/abs/1111.6983

 \bibitem {lev12} M.Sh. Levin,
  Morphological methods for design of modular systems.
 Elect. Preprint,
 (2012) 20 pp.
   http://arxiv.org/abs/1201.1712


 \bibitem {levfim10} M.Sh. Levin, A.V. Fimin,
 Configuration of alarm wireless sensor element.
 In: 2nd Int. Congress on Ultra Modern Telecomm.
 \& Control Syst. and Workshops ICUMT-2010,
 Moscow, (2010) 924-928.

 \bibitem {mcd82} J. McDermott,
 R1: a rule-based configurer of computer systems.
  Artificial Intelligence,  19(2) (1982) 39--88.

 \bibitem {pira05} S. Piramuthu,
 Knowledge-based framework for automated dynamic supply chain
 configuration.
 Eur. J. of Oper. Res., 165 (2005) 219-230.

   \bibitem {qi01} H. Qi, S.S. Iyengar, K. Chakrabarty,
 Distributed sensor networks - a review of recent research,
  J. of the Franklin Institute, 338(6) (2001) 655--668.

  \bibitem {rod05} M.A. Rodriguez, M.C. Jarur,
 A genetic algorithm for searching spatial configurations.
  IEEE Trans. on Evolutionary Computation,
   9(3) (2005)  252--270.

 \bibitem {roy96} B. Roy,
  Multicriteria Methodology for Decision Aiding,
 Kluwer Academic Publishers, Dordrecht, 1996.

  \bibitem {sabin98} D. Sabin, R. Weigel,
  Product configuration frameworks - a survey.
   IEEE Intell. Syst. \& Their Appl.,
    13(4) (1998) 42--49.

 \bibitem {schmidt00} L.C. Schmidt, H. Shetty, S.C. Chase,
 A graph grammar approach for structure synthesis of mechanisms.
  J. Mechanical Design,   122(4) (2000) 371--376.

 \bibitem {smirnov04} A. Smirnov, L. Sheremetov, N. Chilov, J.R. Cortes,
 Soft-computing technologies for configuration of cooperative
 supply chain.
  Applied Soft Computing, 4(1) (2004) 87-107.

 \bibitem {soh07} K. Sohraby, D. Minoli, T. Znati,
  Wireless Sensor Networks: Technology, Protocols, and Applications.
 J.Wiley \& Sons,  New York, 2007.

  \bibitem {poladian06}  J.P. Sousa, V. Poladian, D. Garlan, B. Schmerl, M. Shaw,
 Task-based adaptation for ubiquitous computing.
  IEEE Trans. on SMC - Part C, 36(3) (2006) 328--340.

 \bibitem {stefik95} M. Stefik,
   Introduction to Knowledge Systems,
   Morgan Kaufmann,
  San Francisco, CA, 1995.

  \bibitem {sugu08} V. Sugumaran,  M. Tanniru, V.C. Storney,
 A knowledge-based framework for extracting components in agile
 systems development.
  Inform. Techn. \& Manag.,  9(1) (2008) 37-53.

  \bibitem {wielinga97} B. Wielinga, G. Schreiber,
 Configuration-design problem solving.
  IEEE Expert: Intell. Syst. and Their Appl.,
   12(2) (1997) 49-56.

 \bibitem {zad07} V. Zadorozhny, P. Chrysanthis, P. Krishnamurthy,
  Network-Aware Wireless Sensor Management.
 Springer, New York, 2007.

  \bibitem {zwi69} F. Zwicky,
 Discovery Invention, Research Through the Morphological Approach.
   McMillan, New York, 1969.

\end{thebibliography}
\end{document}